\documentclass{ws-procs9x6}

\def\ga{\gamma}

\def\ka{\kappa}
\def\la{\lambda}

\def\ps{\psi}

\def\cL{{\cal L}}

\def\mn{{\mu\nu}}

\def\fr#1#2{{{#1} \over {#2}}}

\def\frac#1#2{{\textstyle{{#1}\over {#2}}}}

\def\etal{{\it et al.}}

\def\pt#1{\phantom{#1}}

\def\sb{\overline{s}}

\def\ab{\overline{a}}
\def\cb{\overline{c}}

\def\lrDmu{{\hskip -3 pt}\stackrel{\leftrightarrow}{D_\mu}{\hskip -2pt}}
\def\ivb#1#2{e^{#1}_{{\pt{#1}}#2}}
\def\uvb#1#2{e^{#1#2}}
\def\vb#1#2{e_{#1}^{{\pt{#1}}#2}}

\newcommand{\beq}{\begin{equation}}
\newcommand{\eeq}{\end{equation}}
\newcommand{\bea}{\begin{eqnarray}}
\newcommand{\eea}{\end{eqnarray}}

\newcommand{\rf}[1]{(\ref{#1})}

\def\etal {{\it et al.}}

\begin{document}

\title{GRAVITY COUPLINGS IN THE\\
 STANDARD-MODEL EXTENSION}

\author{QUENTIN G.\ BAILEY}

\address{Physics Department,
Embry-Riddle Aeronautical University\\
3700 Willow Creek Road,
Prescott, AZ 86301, USA\\
Email: baileyq@erau.edu} 

\begin{abstract}
The Standard-Model Extension (SME) is an action-based expansion 
describing general Lorentz violation for known matter and fields, 
including gravity.
In this talk, 
I will discuss the Lorentz-violating gravity couplings in the SME.
Toy models that match the SME expansion, 
including vector and two-tensor models, 
are reviewed.
Finally I discuss the status of 
experiments and observations probing gravity coefficients 
for Lorentz violation.
\end{abstract}

\bodymatter

\section{Introduction}

General Relativity (GR) and the Standard Model of particle physics
provide a comprehensive and successful description of nature.
Nonetheless, 
it is expected that an underlying unified description 
containing both theories as limiting cases exists, 
presumably at the Planck scale.  
So far, 
such a complete unified theory remains largely unknown.
Moreover, 
direct measurements at the Planck scale are infeasible at present
so experimental clues about this underlying theory are sparse.

One promising approach is to study suppressed effects 
that may come from the underlying theory. 
An intriguing class of signals that are potentially detectable 
in modern sensitive experiments are minuscule violations of 
local Lorentz symmetry.\cite{reviews} 
A comprehensive effective field theory framework exists called
the Standard-Model Extension (SME)\cite{sme,k04} that 
describes the observable signals of Lorentz violation.
In this framework, 
the degree of Lorentz violation for each type of matter 
or field is controlled by its coefficients for Lorentz violation, 
which vanish when Lorentz symmetry holds.

So far, 
theoretical and experimental work on the SME has 
mostly involved the Minkowski-spacetime limit.\cite{datatables}
Lorentz violation in the gravitational sector
remains comparatively unexplored.
In this talk, 
we focus on two basic types of Lorentz violation involving
gravity: pure-gravity couplings and matter-gravity couplings.
For a more detailed discussion of these topics, 
the reader is referred to Refs.\ \refcite{qbk06,tk09,tk10}.

\section{Theory}

The SME with both gravitational and nongravitational couplings 
was presented in the context of a Riemann-Cartan spacetime
in Ref.\ \refcite{k04}.
In the matter sector of the SME,
Dirac spinor fields can be used for describing the 
matter-gravity couplings that are 
expected to dominate in many experimental scenarios. 
In  this limit the Lagrange density takes the form
\beq
\cL_m = \frac 12 i e \ivb{\mu}{a} {\overline \ps} 
(\ga^a 
- c_{\nu\la} \uvb{\la}{a}\ivb{\nu}{b}\ga^b+...) 
\lrDmu \ps 
-e {\overline \ps} 
(m+a_\mu \ivb{\mu}{a} \ga^a +... ) \ps+...,
\label{matter}
\eeq
where the ellipses represent additional coefficients in the SME
omitted here for simplicity.
The standard vierbein ($\vb{\mu}{a}$) formalism is used 
to incorporate the spinor fields $\ps$ and the gamma matrices 
$\ga^a$ into the tangent space at each point in the spacetime.
Both the spacetime connection and the $U(1)$ connection 
are included in the covariant derivative.
The quantities $c_\mn$ and $a_\mu$ are species-dependent 
coefficients for Lorentz violation. 

In the Riemann-spacetime limit,
the Lagrange density for the pure-gravity sector 
of the SME takes the form
\bea
\cL_g &=& \fr {1}{2\ka} e[(1-u)R + s^\mn R^T_\mn 
+ t^{\ka\la\mu\nu} C_{\ka\la\mu\nu}]+ \cL^\prime.
\label{gravity}
\eea
The 20 coefficients for Lorentz violation 
$u$, 
$s^\mn$, 
and $t^{\ka\la\mu\nu}$ control the leading Lorentz-violating 
gravitational couplings in this expression. 
The curvature tensors appearing are the Ricci scalar $R$, 
the trace-free Ricci tensor $R^T_\mn$ and the Weyl conformal tensor   
$C_{\ka\la\mu\nu}$. 
By convention $\ka=8\pi G$, 
where $G$ is Newton's gravitational constant.  
The additional term $\cL^\prime$ contains the
matter sector and possible dynamical terms
governing the $20$ coefficients.
General coordinate invariance is maintained by
the SME action while
local Lorentz transformations
and diffeomorphisms of the matter 
and gravitational fields are not respected 
by the SME action when $\cL^\prime=0$.

Some geometric constraints arise
when Lorentz violation is introduced in the context
of Riemann-Cartan geometry.
When the coefficients for Lorentz violation 
in the matter and gravity sectors are nondynamical or prescribed 
functions this generally conflicts with the Bianchi identities.
However, 
when the coefficients arise through a dynamical process, 
conflicts with the geometry are avoided.\cite{k04}
This includes spontaneous Lorentz-symmetry breaking scenarios.
The coefficients for Lorentz violation are treated as
arising from spontaneous Lorentz-symmetry breaking
in the approach of Refs.\ \refcite{qbk06,tk09,tk10}.

It is generally a challenging task to study the gravitational
effects in Eqs.\ \rf{matter} and \rf{gravity} in a generic, 
model-independent way. 
It turns out that some simplifications to the analysis
arise in the linearized gravity regime
and it is then possible to extract effective linearized Einstein
equations and modified equations of motion for matter,
under certain assumptions on the dynamics
of the coefficients for Lorentz violation.
These equations then only involve the vacuum expectation
values of the coefficients for Lorentz violation
which are denoted as $\ab_\mu$, 
$\cb_\mn$, 
and $\sb_\mn$. 
Due to species dependence, 
$\ab_\mu$ and $\cb_\mn$ contain $12$ and $27$ independent 
coefficients for ordinary matter, 
respectively.
In the pure-gravity sector, 
only the $9$ species-independent 
$\sb_\mn$ coefficients appear in the linearized gravity limit.

In the post-newtonian limit, 
the metric for the SME can be constructed
from the effective Einstein equations.
An interesting feature arises that
terms in the metric acquire a novel species dependence from 
the $\ab_\mu$ and $\cb_\mn$ coefficients. 
One can also attempt to match to the standard Parametrized 
Post-Newtonian (PPN) formalism.\cite{cmw} 
This involves constraining $\sb_\mn$ 
to an isotropic form in a special coordinate system 
with only one independent coefficient $\sb_{00}$.
Therefore there is a partially overlapping relationship
between the two approaches,
and the SME offers new types of signals for 
gravitational tests.\cite{qbk06} 

\section{Toy models}

Several models of spontaneous Lorentz-symmetry breaking exist
that have a connection to the general formalism described above.
The simplest types of models involve a dynamical vector field
$B_\mu$ that acquires a vacuum expectation value $b_\mu$ 
via a potential term in the lagrangian, which are generically
called bumblebee models.  
Bumblebee models can produce effective $s_\mn$, 
$c_\mn$, 
and $a_\mu$ terms.\cite{qbk06,tk10}
Another interesting class of models 
involves an antisymmetric two-tensor field $B_\mn$.\cite{phon}
The modes appearing in a minimal version of 
these models can include a scalar as well as nondynamical massive modes, 
in addition to producing a background vacuum expectation value $b_\mn$.
Furthermore, 
flat spacetime theories with a self interacting $B_\mn$ field may 
only be stable and renormalizable when the potential admits a
nontrivial minima $b_\mn$, 
thus spontaneously breaking Lorentz symmetry.
When nonminimal couplings to gravity are included, 
these models can also produce effective $\sb_\mn$ coefficients.
Furthermore, 
it can be shown that these effective $\sb_\mn$ coefficients 
cannot be reduced to an isotropic form, 
and so lie outside of PPN analysis.

\section{Matter-gravity tests}

The dominant effects from the coefficients $\ab_\mu$ and $\cb_\mn$
are modified equations of motion for bodies interacting 
gravitationally.
Due to the particle species dependence of these coefficients, 
the motion of a macroscopic body in a gravitational field 
will depend on its internal composition.
This constitutes a violation of the weak equivalence
principle (WEP), 
so the coefficients control WEP violation as well.\cite{tk10}
Existing and proposed tests that can probe these coefficients
include ground-based gravimeter, 
atom interferometry, 
and WEP experiments.
Also of interest are lunar and satellite 
laser ranging observations as
well as measurements of the perihelion precession 
of the planets.

Among the most sensitive tests are proposed satellite
missions designed to test WEP in a microgravity environment.
The observable of interest
for these tests is the relative acceleration
of two test bodies of different composition.
When the relative acceleration is calculated in the satellite 
reference frame in the presence of SME coefficients $\ab_\mu$ and $\cb_\mn$, 
some interesting time-dependent effects arise.
The standard reference frame for reporting coefficient measurements
in the SME is the Sun-centered celestial equatorial reference frame
or SCF for short.\cite{km02}
Upon relating the satellite frame coefficients to the SCF, 
oscillations in the relative acceleration occur at a number of 
different frequencies including multiples and combinations
of the satellite's orbital and rotational frequencies, 
as well as the Earth's orbital frequency. 
This time dependence allows for the extraction
of Lorentz-violating amplitudes independently of 
the standard tidal effects.
Future space-based WEP tests offer sensitivities 
ranging from $10^{-7}$ GeV to $10^{-16}$ GeV for $\ab_\mu$
and $10^{-9}$ to $10^{-16}$ for $\cb_\mn$.
Of particular interest are the STEP,\cite{step}
MicroSCOPE,\cite{micro} and Galileo Galilei\cite{gg} experiments.

\section{Pure-gravity sector tests}

The primary effects due to the nine 
coefficients $\sb_\mn$ in the pure-gravity sector 
of the SME can be obtained from the post-newtonian metric 
and the standard geodesic equation for test bodies.\cite{qbk06} 
Tests potentially probing these coefficients 
include Earth-laboratory tests with gravimeters, 
torsion pendula, 
and short-range gravity experiments.
Space-based tests include lunar and satellite laser ranging,
studies of the secular precession of orbital elements
in the solar system and with binary pulsars, 
and orbiting gyroscope experiments. 

Some analysis placing constraints on the $\sb_\mn$ coefficients
has already been reported.
Using lunar laser ranging data spanning over three decades, 
Battat, Chandler, and Stubbs placed constraints on $6$ combinations of the 
$\sb_\mn$ coefficients at levels of $10^{-7}$ to $10^{-10}$.\cite{ba07}
The modified local acceleration on the Earth's surface was 
measured by M\"uller {\it et al.} using an atom 
interferometric gravimeter, 
resulting in $7$ constraints on the $\sb_\mn$ coefficients 
at the level of $10^{-6}$ to $10^{-9}$.\cite{atom}

Recently, 
the modifications of the classic GR time-delay effect
due to the $\sb_\mn$ coefficients were studied.\cite{qb09}
By studying light propagation with 
the post-newtonian metric modified by the
$\sb_\mn$ coefficients, 
the correction to the light travel time for a signal 
passing near a mass $M$ has been obtained.
Time-delay tests could be particularly useful for 
constraining the isotropic $\sb_{TT}$ coefficient, 
and future tests could yield competitive sensitivities 
to the $\sb_{JK}$ coefficients.
Measurements of $\sb_\mn$ coefficients could be obtained
by using data from time-delay tests such as 
Cassini and BepiColombo.\cite{tdtests}
Also under study are modifications from $\sb_\mn$ coefficients 
to the classic light-bending formula in GR.\cite{rt10}


\begin{thebibliography}{xx}

\bibitem{reviews}
R.\ Bluhm, Lect.\ Notes Phys.\ {\bf 702}, 191 (2006).

\bibitem{sme}
D.\ Colladay and V.A.\ Kosteleck\'y,
Phys.\ Rev.\ D {\bf 55}, 6760 (1997);
Phys.\ Rev.\ D {\bf 58}, 116002 (1998).

\bibitem{k04}
V.A.\ Kosteleck\'y,
Phys.\ Rev.\ D {\bf 69}, 105009 (2004).

\bibitem{datatables}
{\it Data Tables for Lorentz and CPT Violation,}
V.A.\ Kosteleck\'y and N.\ Russell,
2010 edition,
arXiv:0801.0287v3.

\bibitem{qbk06}
Q.G.\ Bailey and V.A.\ Kosteleck\'y,
Phys.\ Rev.\ D {\bf 74}, 045001 (2006).

\bibitem{tk09}
V.A.\ Kosteleck\'y and J.D.\ Tasson,
Phys.\ Rev.\ Lett.\ {\bf 102}, 010402 (2009).

\bibitem{tk10}
V.A.\ Kosteleck\'y and J.D.\ Tasson,
arXiv:1006.4106v1.

\bibitem{cmw}
C.M.\ Will,
Living Rev.\ Relativity {\bf 9}, 3 (2006).

\bibitem{phon}
B.\ Altschul \etal, Phys.\ Rev.\ D {\bf 81}, 065028 (2010).

\bibitem{km02}
V.A.\ Kosteleck\'y and M.\ Mewes,
Phys.\ Rev.\ D {\bf 66}, 056005 (2002).

\bibitem{step}
T.J.\ Sumner \etal, Adv.\ Space Res. {\bf 39}, 254 (2007); 
P.\ Worden, these proceedings.

\bibitem{micro}
P.\ Touboul \etal, 
Comptes Rendus de l'Acad\'emie des Sciences, Series IV, 
{\bf 4}, 1271 (2001).

\bibitem{gg}
A.M. Nobili \etal, Exp.\ Astron.\ {\bf 23}, 689 (2009).

\bibitem{ba07}
J.B.R.\ Battat \etal, 
Phys.\ Rev.\ Lett.\ {\bf 99}, 241103 (2007).

\bibitem{atom}
H.\ M\"uller \etal, 
Phys.\ Rev.\ Lett.\ {\bf 100}, 031101 (2008);
K.-Y.\ Chung \etal,
Phys.\ Rev.\ D {\bf 80}, 016002 (2009).

\bibitem{qb09}
Q.G.\ Bailey,  
Phys. Rev. D {\bf 80}, 044004 (2009). 

\bibitem{tdtests}
B.\ Bertotti, L.\ Iess, and P.\ Tortura,
Nature {\bf 425}, 374 (2003);
L.\ Iess and S.\ Asmar, 
Int.\ J.\ Mod.\ Phys.\ D {\bf 16}, 2191 (2007).

\bibitem{rt10}
R.\ Tso, these proceedings.

\end{thebibliography}
\end{document}